\documentclass[pra,twocolumn,10pt,superscriptaddress]{revtex4-1}  %
\usepackage{graphicx,epsfig,epstopdf}
\usepackage[dvipsnames]{xcolor}
\usepackage{amsmath}
\usepackage{subfigure}
\usepackage{mathrsfs}
\usepackage{bm}
\usepackage{graphicx}
\usepackage{times}

\begin{document}

\title{Nonadiabatic geometric quantum computation with optimal control on superconducting circuits}

\author{Jing Xu}
\thanks{These authors contributed equally to this work.}

\author{Sai Li}
\thanks{These authors contributed equally to this work.}

\author{Tao Chen}
\affiliation{Guangdong Provincial Key Laboratory of Quantum Engineering and Quantum Materials, 
and School of Physics\\ and Telecommunication Engineering, South China Normal University, Guangzhou 510006, China}

\author{Zheng-Yuan Xue}\email{zyxue83@163.com}
\affiliation{Guangdong Provincial Key Laboratory of Quantum Engineering and Quantum Materials, 
and School of Physics\\ and Telecommunication Engineering, South China Normal University, Guangzhou 510006, China}
\affiliation{Frontier Research Institute for Physics, South China Normal University, Guangzhou 510006, China}
	
\date{\today}

\begin{abstract}
  Quantum gates, which are the essential building blocks of quantum computers, are very fragile. Thus, to realize robust quantum gates with high fidelity is the ultimate goal of quantum manipulation. Here, we propose a nonadiabatic geometric quantum computation scheme on superconducting circuits to engineer arbitrary quantum gates, which share both the robust merit of geometric phases and the capacity to combine with optimal control technique to further enhance the gate robustness. Specifically, in our proposal, arbitrary geometric single-qubit gates can be realized on a transmon qubit, by a resonant microwave field driving, with both the amplitude and phase of the driving being time-dependent. Meanwhile, nontrivial two-qubit geometric gates can be implemented by two capacitively coupled transmon qubits, with one of the transmon qubits' frequency being modulated to obtain effective resonant coupling between them.  Therefore, our scheme provides a promising step towards fault-tolerant solid-state quantum computation.
\end{abstract}

\maketitle

\section{Introduction}

Recently, constructing the quantum computer based on the quantum mechanical theory, is highly desired, to deal with hard problems. However, quantum systems will inevitably interact with their surrounding environment. On the other hand, the precise quantum manipulation of a quantum system is limited by the precision of controlling the driving fields. Thus,  fast and robust quantum manipulation is highly desired. To construct a fault-tolerant quantum computer, topological quantum computation strategy is one of the most exciting advances. However, the realization of an elementary quantum gate there is still an experimental difficulty currently. Notably, geometric phases \cite{GP1,GP2,GP3} possess the intrinsic character of noise-resilience against certain local noises \cite{AN1,AN2,AN3,AN4}, and thus can naturally be used to construct robust quantum gates for constructing a fault-tolerant quantum computer.

Previously, geometric quantum computation (GQC) has been proposed based on adiabatic evolutions. 
However, adiabatic evolution requires long running time such that quantum state can be ruined by the decoherence effect. To overcome this problem, GQC with nonadiabatic evolutions has been proposed to achieve high-fidelity quantum gates based on both Abelian \cite{NA1,NA2,NA3,NA4,NA42,NA5} with experimental demonstrations \cite{exp1,exp2,exp3, exp4, chu2019,xuy2019,NA6} and non-Abelian geometric phases  \cite{NNA1,NNA2}. Unfortunately, the existence of systematic errors will devastate the advantage of the robustness of geometric quantum gates \cite{zheng,jing}. Recently,   theoretical \cite{ONN1,ONN2} and experimental works \cite{yan2019, aimz2019} have been proposed to further enhance the robustness of nonadiabatic non-Abelian geometric quantum gates against the control errors, based on three-level systems, by combining the gate operations with optimal control technique (OCT) \cite{OC1,OC2,OC3,OC4,OC5,OC6}. However, compared to the non-Abelian case, quantum gates induced from Abelian geometric phases based on two levels are easier to be realized, and the required two-qubit interaction is experimentally accessible.

Therefore, we here propose a fast GQC scheme that can be compatible with OCT on superconducting circuits, to further improve the robustness of the implemented quantum gates against control errors of the driving fields. Superconducting circuits \cite{SC1,SC2,SC3,SC4} have shown the unique merits of the large-scale integrability and flexibility of operations \cite{fp1, fp2, fp3}, and thus are treated as one of the promising platforms for the physical implementation of scalable quantum computation. Meanwhile, a superconducting transmon device  can be easily addressed to be a two-level system, i.e., the ground and first excited states $\{|0\rangle,|1\rangle\}$, which can serve as a qubit and operated by a resonant driving microwave field. Thus, arbitrary geometric single-qubit gates in our scheme can be accurately achieved after canceling the leakage to the higher excited states, mainly the second excited state $|2\rangle$,  by combing with the DRAG correction \cite{DR1,DR2,DR3}.
In addition, a recent experiment \cite{chu2019} shows that the time-dependent effective resonant coupling can be induced in a two coupled superconducting qubits system, and thus our nontrivial geometric two-qubit control-phase gates can also be resonantly realized \cite{CP1,CP2} in a simple experimental setup. Furthermore, by combining with OCT, the robustness of the implemented geometric gates against the static systematic error can be greatly enhanced.

\section{Geometric single-qubit gates}

In this section, we first explain how to construct arbitrary single-qubit gates on a transmon qubit, based on Abelian geometric phases induced from cyclic evolutions. Then, geometric rotations around the X and Z axes are discussed in detail by faithful numerical simulations with the DRAG correction. Finally, we show that the gate robustness can be further enhanced by combining with OCT.

\begin{figure}[tbp]
  \centering
  \includegraphics[width=0.9\linewidth]{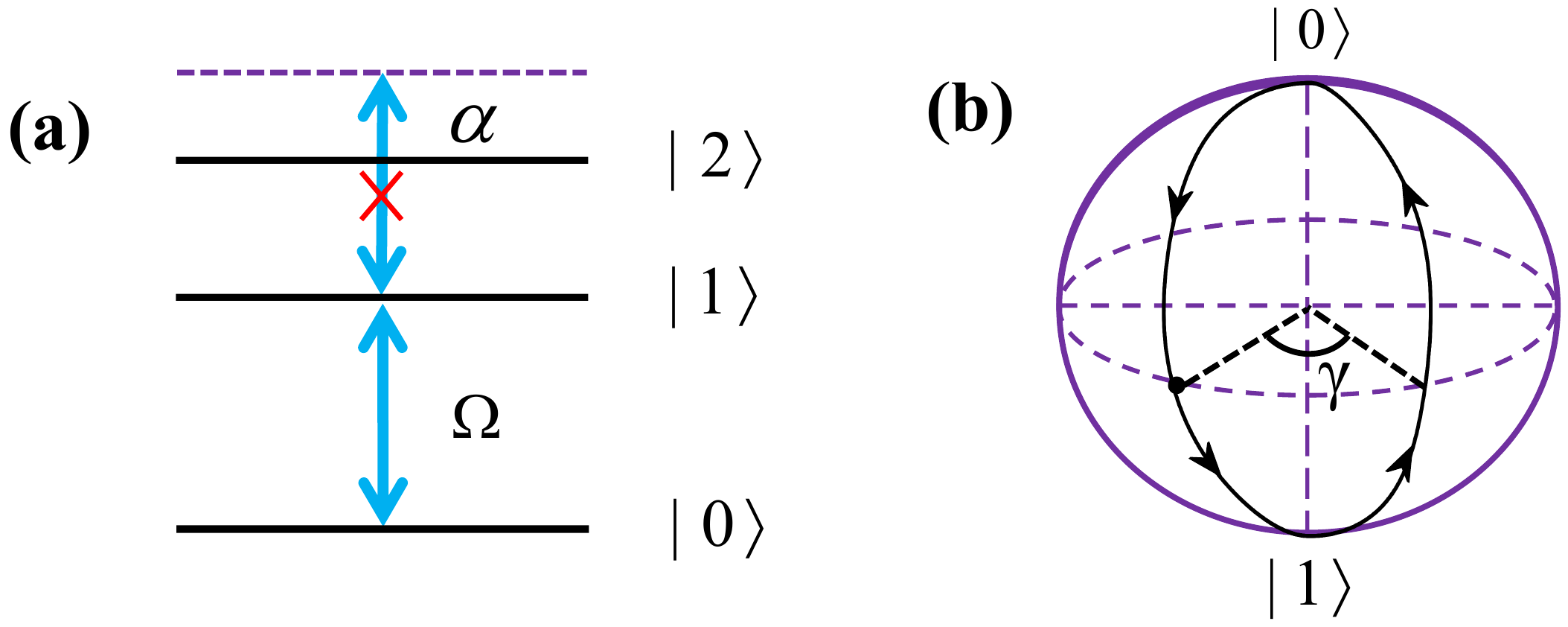}
  \caption{Illustration of our single-qubit geometric quantum gates. (a) The qubit states are resonantly driven to realize arbitrary single-qubit gates, while the driving field will also simultaneously introduce unwanted dispersive transitions to the higher energy states. (b) Geometric illustration of the proposed single-qubit gate on a Bloch sphere.}\label{Figure1}
\end{figure}

\subsection{Construction of the gates}

We consider the construction of arbitrary single-qubit geometric gates in the computation basis $S_1=\{|0\rangle, |1\rangle\}$. For a driving Hamiltonian $H_d ( t )$ in the $S_1$, assuming $\hbar = 1$ hereafter, its dynamic evolution is  governed by the time-dependent Schr\"{o}dinger equation of
\begin{equation}\label{TDSE}
	\begin{aligned}
	i\frac{\partial}{\partial t}|\psi(t)\rangle&=H_{d}(t)|\psi(t)\rangle,
	\end{aligned}
\end{equation}
where
$$|\psi(t)\rangle=e^{-i{f(t) \over 2}}
\left[\cos\frac{\chi(t)}{2}e^{-i{\beta(t)\over 2}}|0\rangle
+\sin \frac{\chi(t)}{2}e^{i{\beta(t) \over 2}}|1\rangle\right]$$
can be generally defined \cite{OC2} by two time-dependent angles $\chi(t)$ and $\beta(t)$, and a parameterized phase $f(t)$. Meanwhile, due to  the linear character of the time-dependent Schr\"{o}dinger equation, the orthogonal evolution state
$$|\psi_{\perp}(t)\rangle=e^{i{f(t)\over 2}}
\left[-\sin \frac{\chi(t)}{2}e^{-i{\beta(t)\over 2}}|0\rangle
+\cos \frac{\chi(t)}{2}e^{i{\beta(t)\over 2}}|1\rangle \right]$$
of $|\psi(t)\rangle$ will also satisfy it.
By modulating the parameters of the driving field, the system undergoes a cyclic evolution, and the initial state $|\psi(0)\rangle$ $(|\psi_\perp(0)\rangle)$ can acquire a global phase $\gamma=[{f(0)-f(\tau)}]/{2}$ $(-\gamma)$ at the final time $\tau$, which consists of a dynamical phase of
\begin{eqnarray}
\label{gamma}
\gamma_D & =&-\int_{0}^{\tau}\langle\psi(t)|H_d(t)|\psi(t)\rangle dt \notag\\
&=& \frac{1}{2}\int_{0}^{\tau}\frac{\dot{\beta}(t) \sin^2 \chi(t)}{\cos \chi(t)}  dt,
\end{eqnarray}
and a geometric phase of
\begin{eqnarray}
\label{gamma}
\gamma_G = i\int_{0}^{\tau}\langle\tilde{\psi}(t)|\dot{\tilde{\psi}}(t) \rangle dt=\frac{1}{2}\int_{0}^{\tau}\dot{\beta}(t)\cos\chi(t) dt,
\end{eqnarray}
where $|{\tilde{\psi}}(t) \rangle=e^{if(t)/2} |{{\psi}}(t) \rangle$. Therefore, by canceling the dynamical phase, i.e., $\gamma_D=0$, in the global phase at the end of the cyclic evolution, we will obtain a pure geometric evolution. Since the evolution here is not governed by the adiabatic condition, the geometric phases will be induced in a  faster way than that of the adiabatic schemes \cite{GP3}. Especially, when the phase in Hamiltonian is a constant, our scheme will reduce to the conventional non-adiabatic schemes \cite{NA3,NA4}. Then, the  final geometric evolution operator in  $S_1$ is
\begin{eqnarray}\label{U1}
	U(\tau) &= &e^{i\gamma}|\psi(0)\rangle \langle \psi(0)| + e^{-i\gamma}|\psi_\perp(0)\rangle \langle \psi_\perp(0)|\notag\\
	 &=&\left(\begin{array}{cc}{\cos \gamma+i \cos \chi_{_0} \sin \gamma} & {i \sin \gamma \sin \chi_{_0} e^{-i \beta_{_0}}} \\ {i \sin \gamma \sin \chi_{_0} e^{i \beta_{_0}}} & {\cos \gamma-i \cos\chi_{_0}\sin \gamma}\end{array}\right) \notag\\
	&=&e^{i\gamma \vec{n} \cdot \vec{\sigma} },
\end{eqnarray}
{where $\chi_{_0}=\chi(0)$, $\beta_{_0}=\beta(0)$.} It is a rotation around the axis of $\vec{n} =(\sin\chi_{_0} \cos\beta_{_0},\sin\chi_{_0} \sin\beta_{_0},\cos\chi_{_0}) $ by an angle $-2\gamma$, from which arbitrary  single-qubit gates can be induced.

\subsection{Gate implementation}

We now proceed to physical implementation of our scheme. As shown in Fig. \ref{Figure1}(a), the two lowest levels $|0\rangle$ and $|1\rangle$ of a single transmon qubit can be resonantly coupled by a microwave field with time-dependent amplitude $\Omega(t)$ and phase $\phi(t)$. Into the interaction picture, neglecting the high-order oscillating terms by the rotating-wave approximation. Then, the Hamiltonian of the system in $S_1$ is
\begin{equation}\label{H0}
\begin{aligned}
H_d(t) = \frac {1} {2}
\left(\begin{array} {cc} {0} & {\Omega(t)e^{i\phi(t)}} \\ { \Omega(t)e^{-i\phi(t)}} & {0} \end{array} \right).
\end{aligned}
\end{equation}
In the following, to let $H_d(t)$ fulfil Eq. (\ref{TDSE}) at any moment, see  Appendix A for details, we can obtain the constraints of parameters as
\begin{subequations}\label{relation1}
\begin{equation}\label{relation1a}
\dot{f}(t)=-\frac{\dot {\beta}(t)}{\cos\chi(t)},
\end{equation}
\begin{equation}\label{relation1b}
\dot{\chi}(t)=-\Omega(t) \sin[\beta(t)+\phi(t)],
\end{equation}
\begin{equation}\label{relation1c}
\dot{\beta}(t)= - \Omega(t) \cot{\chi}(t)\cos{[\beta(t)+\phi(t)]}.
\end{equation}
\end{subequations}
Thus, we find that a target evolution path of the evolution state $|\psi(t)\rangle$ $(|\psi_\perp(t)\rangle)$ can be determined by  designing the parameters $\Omega(t)$ and $\phi(t)$ of the microwave field. For the purpose of constructing arbitrary single-qubit geometric gates, we set a single-loop evolution path by defining the evolution parameters $\chi(t)$ and $\beta(t)$ to fulfill a cyclic evolution, as illustrated in Fig. \ref{Figure1}(b). Thus we can inversely determine the parameters $\Omega(t)$ and $\phi(t)$, i.e.,
\begin{equation}\label{Jp}
\begin{aligned}
	\Omega(t) &= -\frac{\dot{\chi}{(t)}}{\sin{(\beta(t)+\phi(t))}}, \\
	\phi(t)  &= \arctan{\left(\frac{\dot{\chi}{(t)}}{\dot{\beta}{(t)}}\cot{\chi(t)}\right)}-\beta(t).	
\end{aligned}
\end{equation}
In addition, we also need to ensure accumulated dynamical phases are zero at the end of cyclic evolution to achieve pure geometric operations.

Specifically, we consider rotations around the X and Z axes as two typical examples in detail. Firstly, to realize the geometric rotation operators around the X axis, we divide a single-loop evolution path into four equal parts, which aims to cancel dynamical phases at the end of cyclic evolution. The shape of $\chi_j(t)$ and initial values of $\beta_j(t)$ in each part are
\begin{equation}\label{casex}
\begin{aligned}
t\in [0,  \tau/4]: \quad &\chi_{1}(t)=\pi [1+ \sin^2( 2\pi t/\tau)]/2,  \\
&\beta_1(0)=0 ,\\
t \in [\tau/4, \tau/2]: \quad &\chi_{2}(t)=\pi [1+ \sin^2( 2\pi t/\tau)]/2, \\
    & \beta_2( \tau/4)=\beta_1( \tau/4) - {\gamma}, \\
t \in [\tau/2,  3\tau/4]: \quad   & \chi_{3}(t)=\pi [1- \sin^2( 2\pi t/\tau)]/2,  \\
    &  \beta_3(\tau/2)=\beta_2(\tau/2),\\
t  \in [3\tau/4, \tau]: \quad  &  \chi_{4}(t)=\pi [1- \sin^2( 2\pi t/\tau)]/2, \\
    & \beta_4(3\tau/4)=\beta_3(3\tau/4) + {\gamma},\\
\end{aligned}
\end{equation}
where the shape of $\beta_j(t)$ is set as $\beta_j(t)=-\int\dot{f_j}{(t)}\cos\chi_j(t) dt$ for the $j$th part with $f_j(t) = \cos2\chi_j(t)/5$.
Therefore, we can obtain the shape of $\Omega(t)$ and $\phi(t)$ in the different evolution parts according to Eq. (\ref{Jp}). Meanwhile, in this setting, the dynamical phase is
\begin{eqnarray}
\label{gamma}
\gamma_D = \frac{1}{2}\sum^4_{j=1}\int_{(j-1)\tau/4}^{j\tau/4}\frac{\dot{\beta_j}(t) \sin^2 \chi_j(t)}{\cos \chi_j(t)}  dt=0,
\end{eqnarray}
and the geometric phase is $\gamma_G=\gamma$ due to the saltation of $\beta(t)$ at the moment of $t=\tau/4$ and $t=3\tau/4$. In this way, the geometric rotation operations $e^{i\gamma\sigma_x}$ can be realized.

Similarly, to realize the geometric rotation operators around the Z axis, we divide a single-loop evolution path into only two equal parts,  which also aims to cancel dynamical phases at the end of cyclic evolution. The shape of $\chi_j(t)$ and initial values of $\beta_j(t)$ in the each parts are
\begin{equation}\label{casez}
\begin{aligned}
  t \in [0, \tau/2]: \quad  & \chi_{1}(t)=\pi\sin^2( \pi t/\tau),\\
  &\beta_1(0)=0,\\
 t \in [\tau/2, \tau]:\quad &   \chi_{2}(t) =\pi\sin^2( \pi t/\tau), \\
& \beta_2(\tau/2)=\beta_1(\tau/2) - \gamma, \\
\end{aligned}
\end{equation}
where the shape of $\beta_j(t)$ is set to be $\beta_j(t)=-\int\dot{f_j}{(t)}\cos\chi_j(t) dt$ for the $j$th part with $f_j(t)=[2\chi_j(t)-\sin2\chi_j(t)]/5$. The shape of $\Omega(t)$ and $\phi(t)$ can also be obtained in the different evolution parts according to Eq. (\ref{Jp}). By only the saltation of $\beta(t)$ at the moment of $t = \tau/2$, the dynamical phase at the end of the cyclic evolution can be eliminated and  the pure geometric phase can be accumulated.
Therefore, the geometric rotations $e^{i\gamma\sigma_z}$ can be realized.

\begin{figure}[tbp]
  \centering
  \includegraphics[width=0.9\linewidth]{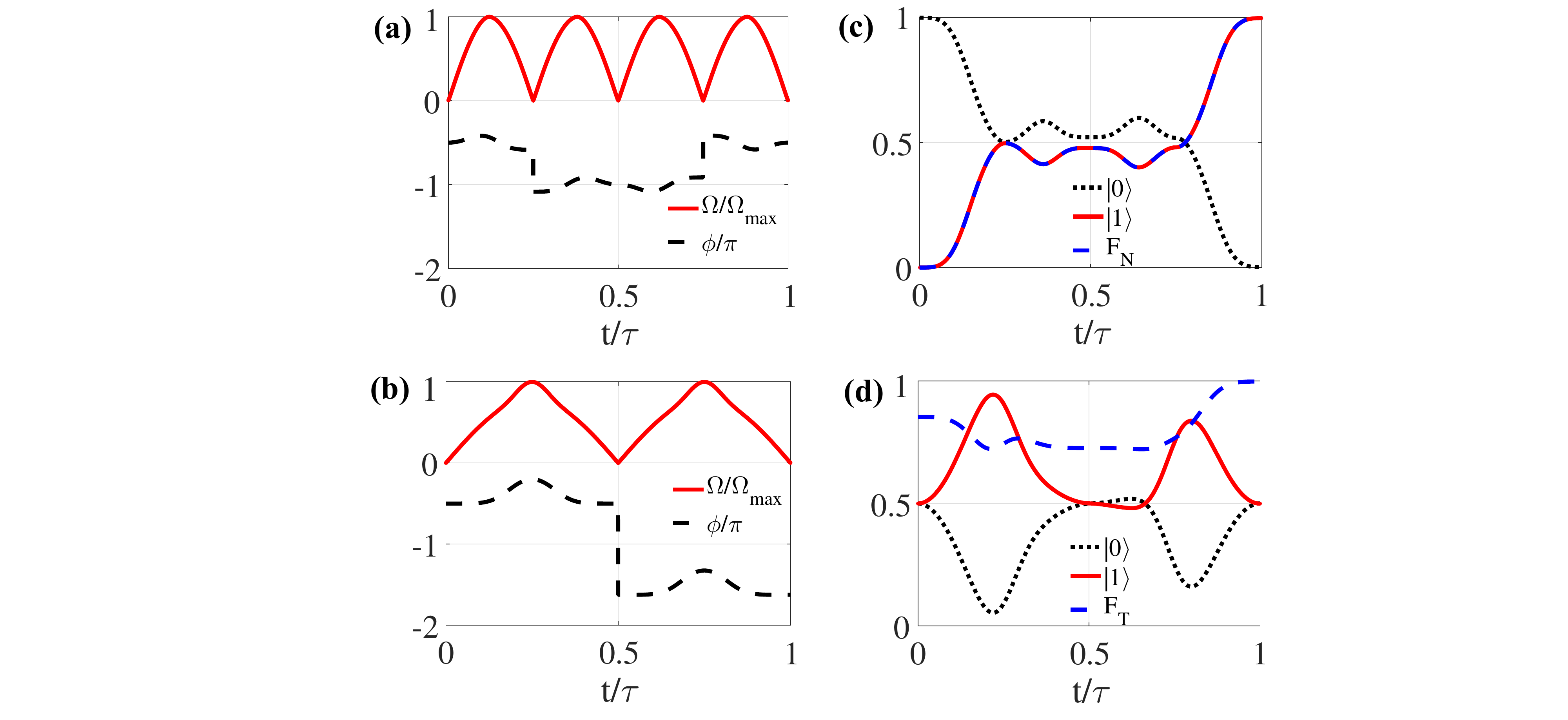}
  \caption{Implementation of single-qubit geometric gates and their performance. The shapes of $\Omega(t)$ and $\phi(t)$ for the NOT and Phase gates are shown in (a) and (b), respectively. The qubit-state population and the state-fidelity dynamics of the NOT and Phase gate operations are shown in (c) and (d), respectively.}\label{Figure2}
\end{figure}

\begin{figure*}[tbp]
  \centering
  \includegraphics[width=0.8\linewidth]{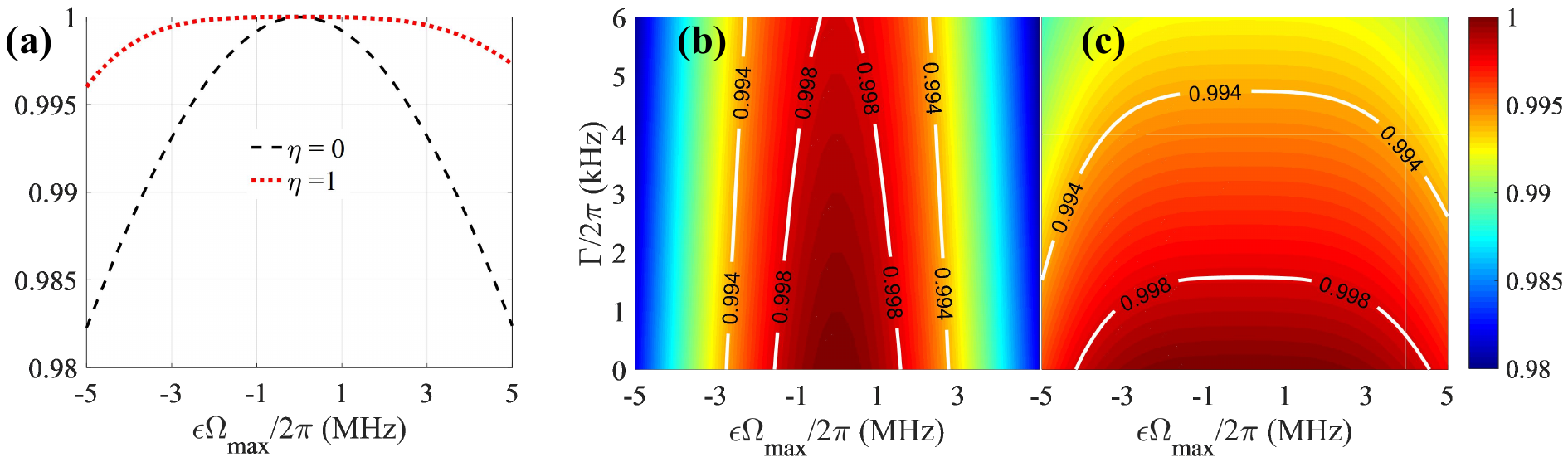}
  \caption{Single-qubit gate performance without ($\eta=0$) and with ($\eta=1$) optimization. (a) The gate fidelity of the Phase gate with different $\eta$ under the systematic error $\epsilon$ without decoherence. The gate fidelity of the Phase gate without and with optimization are shown in (b) and (c), respectively, under both the systematic error $\epsilon\Omega_{\textrm{max}}$ and a uniform decoherence rate $\Gamma$.}\label{Figure3}
\end{figure*}

\subsection{Gate performance}

However, in the practical physical implementation, the effect of decoherence is a non-negligible factor to measure gate performance. Notably, due to the weak anharmonicity of the transmon qubit, here, the DRAG correction \cite{DR1,DR2,DR3} is also introduced to suppress the leakage error beyond the qubit basis. Considering all effects of decoherence and dominant counter-rotating terms, we here use the Lindblad master equation
\begin{equation}
\dot{\rho}_{1}=i \left[\rho_{1}, H_{d}(t)+H_{\textrm{leak}}(t)\right]+ \left[\Gamma_{1} \mathcal{L}\left(\sigma_{1}\right)+\Gamma_{2}\mathcal{L}\left(\sigma_2\right)\right],
\end{equation}
with
\begin{equation}
H_{\textrm{leak}}(t)=-\alpha |2\rangle\langle 2|+\left[\sqrt{2}\Omega(t)e^{i\phi(t)}|1\rangle \langle 2|+\mathrm{H.c.}\right],
\end{equation}
to evaluate the performance of the implemented single-qubit gates, where $\rho_{1}$ is the density matrix of the considered system and $\mathcal{L}(\mathcal{A})= \mathcal{A} \rho_{1} \mathcal{A}^{\dagger} -\mathcal{A}^{\dagger} \mathcal{A} \rho_{1}/2-\rho_{1} \mathcal{A}^{\dagger} \mathcal{A}/2$ is the Lindblad operator $\mathcal{A}$ with $\sigma_{1}=|0\rangle \langle 1| + \sqrt{2}|1\rangle\langle2|$ and $\sigma_2=| 1\rangle \langle 1| + 2| 2\rangle \langle 2|$, and $\Gamma_{1}$ and $\Gamma_{2}$ are the decay and dephasing rates of the transmon qubit, respectively. We consider the case of $\Gamma_{1} = \Gamma_{2}= \Gamma = 2\pi\times 2 $ kHz \cite{SC4}, which is easily accessible with current experimental technologies. The anharmonicity of the transmon is set to be $\alpha = 2\pi \times 300$ MHz, and the maximum amplitude $\Omega_{\textrm{max}} = 2\pi\times 16$ MHz. We next take the NOT ($N$) and Phase ($T$) gates as two typical examples, which can be realized by setting $\chi_1(0)=\pi/2$, $\gamma=\pi/2$ and $\chi_1(0)=0$, $\gamma=-\pi/8$ with the same $\beta_1(0)=0$, respectively.  Under maximum amplitude $\Omega_{\textrm{max}} = 2\pi\times 16$ MHz, the cyclic evolution time $\tau$ is about 102 ns for the NOT gate and 125 ns for the Phase gate. The corresponding shapes of $\Omega(t)$ and $\phi(t)$ for the NOT and Phase gates are shown in Figs. \ref{Figure2}(a) and \ref{Figure2}(b), respectively. Assuming the initial states of quantum system are $|\Phi(0)\rangle_N =|0\rangle$ and $|\Phi(0)\rangle_T =(|0\rangle+|1\rangle)/\sqrt{2}$ for the cases of the NOT and Phase gates, these geometric gates can be evaluated by using the state fidelity defined by $F_{N/T}=_{N/T}\langle\Phi(\tau)|\rho_1|\Phi(\tau)\rangle_{N/T}$ with $|\Phi(\tau)\rangle_N=|1\rangle$ and $|\Phi(\tau)\rangle_T=(|0\rangle+e^{i\pi/4}|1\rangle)/\sqrt{2}$ being the corresponding ideal final states. The obtained fidelities are as high as $F _{N}= 99.87\%$ and $F _{T}= 99.80\%$, as shown in Figs. \ref{Figure2}(c) and \ref{Figure2}(d), respectively. In addition, for the general initial state $|\Phi_1(0)\rangle=\cos\theta|0\rangle+\sin\theta|1\rangle$, the NOT and Phase gates should result in the ideal final states $|\Phi(\tau)\rangle_{N}=\sin\theta|0\rangle+\cos\theta|1\rangle$ and $|\Phi(\tau)\rangle_{T}=\cos\theta
|0\rangle+e^{i\pi/4}\sin\theta|1\rangle$. To fully evaluate the gate performance, we define gate fidelity as $F^G_{N/T}=(\frac{1}{2\pi}){\int^{2\pi}_0}_{N/T}\langle\Phi(\tau)|\rho_1|\Phi(\tau)\rangle_{N/H} d\theta$ with the integration numerically performed for 1001 input states with $\theta$ being uniformly distributed over $[0,2\pi]$. We find that the gate fidelities of the NOT and Phase gates can, respectively, reach $F^G _{N}= 99.87\%$ and $F^G _{T}= 99.84\%$.

\subsection{Optimal control}
In the above, we have presented and numerically demonstrated our scheme for implementing geometric single-qubit gates, where the evolution path are designed by choosing proper parameters in Hamiltonian $H_d(t)$, which are inversely engineered. Here, we proceed to design the evolution path by combining it with OCT \cite{OC1,OC2}, to further enhance the robustness of our scheme against systematic error. Specifically, in the case of the geometric rotation operators around the Z axis, we consider the existence of the static systematic error, i.e. $\Omega(t)\rightarrow(1+\epsilon)\Omega(t)$.
Due to the symmetry of the evolution path of the considered geometric rotations, we take the first path $[0, \tau/2]$ to evaluate the  gate robustness,  which can be calculated by the perturbation theory with probability amplitude  $P$ defined  as
\begin{equation}
	\begin{aligned}
P=|\langle \psi({\tau}/{2})|\psi_\epsilon({\tau}/{2})\rangle|^2=1+\tilde{O}_{1}+\tilde{O}_{2}+\cdots,
	\end{aligned}
\end{equation}
where $|\psi_{\epsilon}(\tau/2) \rangle$ is the state with the systematic error, and $\tilde{O}_n$ denotes the term of the perturbation at the $n$th order. Ignoring the high-order terms, we next just consider $P$ to the second order, i.e., $P_2 = 1+\tilde{O}_{1}+\tilde{O}_{2}$, which is calculated to be
\begin{equation}\label{Qs}
\tilde{O}_{1} =0, \quad \tilde{O}_{2}=-\epsilon^2 \left|\int^{\chi_\frac{\tau}{2}}_{\chi_0} e^{-if}\sin^2{\chi} d\chi \right|^2.
\end{equation}
Defining $f(\chi) =\eta[2\chi - \sin{(2\chi)]}$, $\beta_1(0)=0$ and $\beta_2(\tau/2)= \beta_1(\tau/2)-\gamma$, resulting in $\tilde{O}_{2}=-\epsilon^2\sin^2\eta\pi/(2\eta)^2$, we can ensure $\tilde{O}_{2}=0$ for $P_2 =1$ by setting $\eta$ to a non-zero integer,  which directly demonstrates that the designed evolution path is combined with OCT. It is important to note that when $\eta=0$, $\tilde{O}_{2}= -\pi^2\epsilon^2/4$, the current implementation will reduce to the previous non-adiabatic schemes \cite{NA1,NA2,NA3,NA4}. In the following numerical simulations, for a fair comparison, all the maximum value of $\Omega(t)$ are set to be $\Omega_{\textrm{max}}= 2\pi\times 16$ MHz. That is the maximum value of the optimized pulse is bounded by $\Omega_{\textrm{max}}$, and thus the improvement of the gate performance can only be attributed to OCT.

For a typical example, we simulate the Phase gate under the effect of the systematic error $\epsilon\Omega_{\textrm{max}}$ in the range of $2\pi\times[-5,5]$ MHz.
In Fig. \ref{Figure3}(a), we plot the gate fidelity as a function of the systematic error $\epsilon\Omega_{\textrm{max}}$ for the cases of $\eta=0$ and $\eta=1$ without decoherence with the cyclic evolution time $\tau$ being 98 ns  and 405 ns, respectively. We find that the robustness of our geometric gates in the case of $\eta=1$ can be significant improved comparing with the case of $\eta=0$ (previous implementations). Meanwhile, as shown in Figs. \ref{Figure3}(b) and \ref{Figure3}(c), considering both the systematic error and the decoherence effects, our geometric gate based on OCT still has the advantage of improving robustness in a certain decoherence range.

\section{Nontrivial geometric two-qubit gates}

\begin{figure}[tbp]
  \centering
  \includegraphics[width=0.9\linewidth]{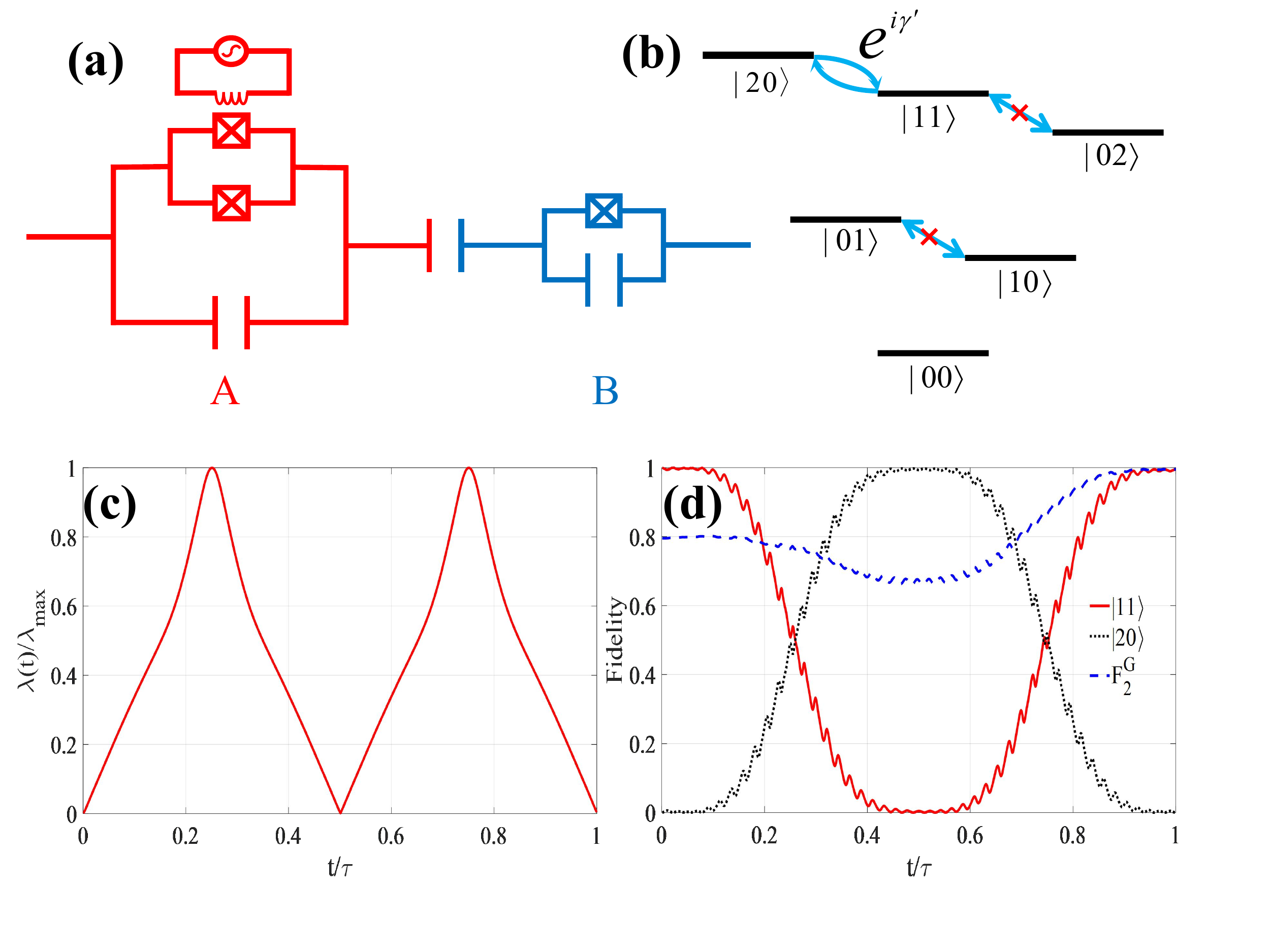}
  \caption{Illustration of the implementation of the two-qubit geometric gates. (a) Two capacitively coupled transmon qubits configuration for non-trivial two-qubit gates, where the frequency of qubit A is ac modulated to induce effective resonant interaction between the two qubits. (b) The coupling structure for the states of the two transmon qubits. (c) the control pulse envelope of $\lambda(t)$. (d) State performance and the gate fidelity of a nontrivial geometric control-phase gate with $\gamma^\prime = \pi/2$.}\label{Figure4}
\end{figure}

In this section, we turn to the implementation of nontrivial two-qubit geometric gates based on two capacitively coupled transmons \cite{CC1,CC2,CC3}, which are respectively labeled by transmons A and B with qubit frequency $\omega_{\textrm{A},\textrm{B}}$ and anharmonicity $\alpha_{\textrm{A},\textrm{B}}$. However, the frequency difference $\Delta=\omega_\textrm{A}-\omega_\textrm{B}$ and coupling strength $g$ between this two transmons A and B are usually fixed and can not be adjustable. Profitably, in a recent experimental setup \cite{chu2019}, as illustrated in in Fig. \ref{Figure4}(a), time-dependent tunable coupling interaction can be realized by introducing a qubit-frequency driving $\zeta(\varepsilon(t))$ on transmon A, which can be experimentally induced by adding a longitudinal field $\varepsilon(t) = \zeta^{-1}(\dot{\mathcal{F}}(t))$, where $\mathcal{F}(t) = \lambda(t)\sin[\nu t + \varphi(t)]$ with $\nu$ and $\varphi(t)$ being the frequency and phase of the longitudinal field, respectively. Then, in the interaction picture, see Appendix B for details, the effective Hamiltonian of the two coupled transmons is
\begin{eqnarray}\label{Hint}
{H}_t(t)=g[|10\rangle_{\textrm{AB}}\langle01|e^{i\Delta t }+ \sqrt{2}|11\rangle_{\textrm{AB}}\langle02|e^{i(\Delta+\alpha_\textrm{B}) t } \notag\\
+\sqrt{2}|20\rangle_{\textrm{AB}}\langle11|e^{i(\Delta-\alpha_\textrm{A}) t }]e^{-i\lambda(t)\sin[\nu t + \varphi(t)]}+\textrm{H.c.}.
\end{eqnarray}
The corresponding coupling configuration of these two coupled transmons is shown in Fig. \ref{Figure4}(b). We consider the case of the resonant interaction in the subspace $\{|11\rangle_{\textrm{AB}}, |20\rangle_{\textrm{AB}}\}$ by choosing the driving frequency $\nu=\Delta-\alpha_\textrm{A}$ with $g\ll\{\nu, \Delta-\nu, \Delta+ \alpha_\textrm{B}-\nu \}$, and then using Jacobi-Anger identity and neglecting the high-order oscillating terms, the obtained effective Hamiltonian can be reduced to
\begin{equation}\label{Heff}
\begin{aligned}
H_{2}(t) = \frac {1} {2} \left(\begin{array} {cc} {0} & {g^\prime(t)e^{i\varphi(t)}} \\ {g^\prime(t)e^{-i \varphi(t)}} & {0} \end{array} \right),
\end{aligned}
\end{equation}
in the two-qubit subspace $\{|11\rangle_{\textrm{AB}}, |20\rangle_{\textrm{AB}}\}$, where $g^\prime(t) = 2\sqrt{2}gJ_1(\lambda(t)) $ is effective time-dependent coupling strength between transmon qubits A and B, with $J_1(\lambda(t))$ being the Bessel function of the first kind.

We note that the Hamiltonian in Eq. (\ref{Heff}) is in the same form as that of the single-qubit case in Eq. (\ref{H0}). Thus, within the two-qubit subspace $\{|00\rangle_{\textrm{AB}}, |01\rangle_{\textrm{AB}}, |10\rangle_{\textrm{AB}}, |11\rangle_{\textrm{AB}}\}$, {we can also use the effective Hamiltonian $H_{2} (t)$ to acquire a pure geometric phase $e^{i\gamma^\prime}$ condition on two-qubit state of $|11\rangle_{\textrm{AB}}$ by a cyclic evolution,  which is just like the way of constructing geometric rotation operations $e^{i\gamma\sigma_z}$.} The resulting nontrivial two-qubit geometric control-phase gates can be obtained as
\begin{equation}\label{U2}
\begin{aligned}
U_2 ( \gamma^\prime ) =  \left(
\begin{array}{cccc}
1 & 0 & 0 & 0 \\
0 & 1 & 0 & 0 \\
0 & 0 & 1 & 0 \\
0 & 0 & 0 & e^{i\gamma^\prime} \\
\end{array}
\right).
\end{aligned}
\end{equation}

Here, we also use the Lindblad master equation to evaluate the nontrivial two-qubit geometric control-phase gates with $\gamma^\prime = \pi/2$ as a typical example. We set the parameters of coupled transmon qubits as $\Delta =2\pi\times 500$ MHz, $\alpha_\textrm{A} =2\pi\times 320$ MHz, $\alpha_\textrm{B} =2\pi\times 300$ MHz, $g = 2\pi\times 5$ MHz and the driving frequency $\nu = \Delta-\alpha_\textrm{A}=2\pi\times 180$ MHz, and the decoherence rate of transmons is the same as the single-qubit case \cite{SC4}. Furthermore, we fix the evolution time $\tau'$ for the two-qubit gate to be $250$ ns under a corresponding coupling strength of $g^\prime_{max}=2\pi\times8$ MHz, and the form of the auxiliary parameters $\chi(t)$, $\beta(t)$ being the same as that of the single-qubit case for the geometric rotation operators around Z axis. In this way, the shape of $g^\prime(t) $ and $\varphi(t)$ can finally be determined. In addition, we can numerically define $\lambda (t) = J_1^{-1}[g^\prime(t)/(2\sqrt2g)]$ as shown in Fig. 4(c), and then use the original Hamiltonian $\mathcal{H}_t(t)$ to faithfully verify our proposal. For the general initial state of the two qubit as $|\Phi_2(0)\rangle = (\cos\vartheta_1 |0\rangle_\textrm{A} + \sin\vartheta_1 |1\rangle_\textrm{A} )\otimes(\cos\vartheta_2 |0\rangle_\textrm{B} + \sin\vartheta_2 |1\rangle_\textrm{B} )$ with $|\Phi_{2}(\tau')\rangle  = U_2(\pi/2)|\Phi_2(0)\rangle$ being the ideal final state, we can define the two-qubit gate fidelity as
\begin{equation}
\label{TG}
F^G_2 = \frac{1}{4\pi^2}\int^{2\pi}_0\int^{2\pi}_0\langle\Phi_{2}(\tau')|\rho_2|\Phi_{2}(\tau')\rangle d\vartheta_1 d\vartheta_2,
\end{equation}
with the integration numerically done for 10001 input states with $\vartheta_1$ and $\vartheta_2$ uniformly distributed over $[0, 2\pi]$. As shown in Fig. \ref{Figure4}(d), we can get the gate fidelity $F^G_2=99.53\%$. Finally, comparing the two-qubit Hamiltonian $H_2(t)$ with the single-qubit Hamiltonian $H_d(t)$, one find that they are in the same form, both with the tunable coupling strength and phase. Thus, the OCT presented in the single-qubit case can be directly incorporated in this two-qubit gate implementation.

\section{CONCLUSION}

In summary, we have proposed a general method to construct fast universal GQC. Then, we physically implement our proposal on superconducting circuits, where arbitrary single-qubit gates are realized by resonant driving on a transmon qubit with a microwave field, and nontrivial two-qubit gates can be implemented by ac driving on one of the transmon qubits, which leads to effectively resonant  coupling between them. Finally, our scheme can combine with OCT to further enhance the gate robustness against the static systematic error. We note that our proposal can be expanded to a two-dimensional capacitively coupled lattice of transmon qubits, and thus provides a promising step towards  fault-tolerant quantum computation on superconducting circuits.

\section*{ACKNOWLEDGMENTS}
This work was supported by the Key-Area Research and Development Program of GuangDong Province (Grant No. 2018B030326001), the National Natural Science Foundation of China (Grant No. 11874156), the National Key R\&D Program of China (Grant No. 2016 YFA0301803).

\appendix
\section{Derivation of Eq.(6)}
Here, we present details of deriving Eq.(6). Firstly, inserting Eq.(5) into Eq.(1) in main text, one obtains
\begin{eqnarray}\label{A1}
&&i\frac{\partial}{\partial t}\left[e^{-if/2}\left(\begin{array} {cc}\cos\frac{\chi}{2}e^{-i\beta/2}\\\sin\frac{\chi}{2}e^{i\beta/2}
\end{array} \right)\right]\notag\\
&&=\frac {1} {2}
\left(\begin{array} {cc} {0} & {\Omega e^{i\phi}} \\ { \Omega e^{-i\phi}} & {0} \end{array} \right)e^{-if/2}\left(\begin{array} {cc}\cos\frac{\chi}{2}e^{-i\beta/2}\\\sin\frac{\chi}{2}e^{i\beta/2}
\end{array} \right),
\end{eqnarray}
then, we can expand it as
\begin{eqnarray}\label{A2}
&&i\left(\begin{array} {cc}
-i\frac{\dot{f}}{2}\cos\frac{\chi}{2}e^{-i\beta/2}
-\frac{\dot{\chi}}{2}\sin\frac{\chi}{2}e^{-i\beta/2}
-i\frac{\dot{\beta}}{2}\cos\frac{\chi}{2}e^{-i\beta/2}\\
-i\frac{\dot{f}}{2}\sin\frac{\chi}{2}e^{i\beta/2}
+\frac{\dot{\chi}}{2}\cos\frac{\chi}{2}e^{i\beta/2}
+i\frac{\dot{\beta}}{2}\sin\frac{\chi}{2}e^{i\beta/2}
\end{array} \right)\notag\\
&&=\frac {1} {2}
\left(\begin{array} {cc}
\Omega\sin\frac{\chi}{2}e^{i(\beta/2+\phi)}\\
\Omega\cos\frac{\chi}{2}e^{-i(\beta/2+\phi)}
\end{array} \right).
\end{eqnarray}
Applying corresponding matrix elements to be equal, we get
\begin{eqnarray}\label{A34}
&&\frac{\dot{f}}{2}\cos\frac{\chi}{2}e^{-i\beta/2}
-i\frac{\dot{\chi}}{2}\sin\frac{\chi}{2}e^{-i\beta/2}
+\frac{\dot{\beta}}{2}\cos\frac{\chi}{2}e^{-i\beta/2}\notag\\
&&=\frac {\Omega} {2}\sin\frac{\chi}{2}e^{i(\beta/2+\phi)},\\
&&\frac{\dot{f}}{2}\sin\frac{\chi}{2}e^{i\beta/2}
+i\frac{\dot{\chi}}{2}\cos\frac{\chi}{2}e^{i\beta/2}
-\frac{\dot{\beta}}{2}\sin\frac{\chi}{2}e^{i\beta/2}\notag\\
&&=\frac {\Omega} {2}\cos\frac{\chi}{2}e^{-i(\beta/2+\phi)}.
\end{eqnarray}
Applying corresponding real part and imaginary part to be equal, we get
\begin{eqnarray}\label{A5678}
&&\frac{\dot{f}}{2}\cos\frac{\chi}{2}
+\frac{\dot{\beta}}{2}\cos\frac{\chi}{2}
=\frac {\Omega} {2}\sin\frac{\chi}{2}\cos{(\beta+\phi)},\\
&&\frac{\dot{\chi}}{2}\sin\frac{\chi}{2}
=-\frac {\Omega} {2}\sin\frac{\chi}{2}\sin{(\beta+\phi)},\\
&&\frac{\dot{f}}{2}\sin\frac{\chi}{2}
-\frac{\dot{\beta}}{2}\sin\frac{\chi}{2}
=\frac {\Omega} {2}\cos\frac{\chi}{2}\cos{(\beta+\phi)},\\
&&i\frac{\dot{\chi}}{2}\cos\frac{\chi}{2}
=-\frac {\Omega} {2}\cos\frac{\chi}{2}\sin{(\beta+\phi)}.
\end{eqnarray}
Combining Eq. (A5) and Eq. (A7), we can get Eqs. (\ref{relation1a}) and (\ref{relation1c}) in the main text.
Similarly, combining Eq. (A6) and Eq. (A8), one can get Eq. (\ref{relation1b}) in the main text.

\section{Effective two-qubit Hamiltonian}
Here, we present the derivation details of the effective two-qubit Hamiltonian in Eq (\ref{Hint}). The coupled system can be described by $H_T(t)= H_f(t)+H_I(t)$, where
$H_f(t)$ is free part and $H_I(t)$ is interaction part. For the free part,
\begin{eqnarray}\label{B1}
H_f(t)&=&[\omega_A+\zeta(\epsilon(t))]|1\rangle_A\langle1|\notag\\
&&+[2\omega_A-\alpha_A +2\zeta(\epsilon(t))]|2\rangle_A\langle2|\notag\\
&&+\omega_B|1\rangle_B\langle1|+(2\omega_
B-\alpha_B)|2\rangle_B\langle2|,
\end{eqnarray}
where $\varepsilon(t) = \zeta^{-1}(\dot{\mathcal{F}}(t))$ with $\mathcal{F}(t) = \lambda(t)\sin[\nu t + \varphi(t)]$. For interaction term,
\begin{eqnarray}\label{B2}
H_I(t)=&&g(|0\rangle_A\langle1|+\sqrt{2}|1\rangle_A\langle2|+\textrm{H.c.})\notag\\
&&\cdot(|0\rangle_B\langle1|+\sqrt{2}|1\rangle_B\langle2|+\textrm{H.c.}).
\end{eqnarray}
Moving to the rotating frame defined by $V=V_1+V_2$, where
\begin{eqnarray}\label{B3}
V_1 =\exp&&\{-i[\omega_A|1\rangle_A\langle1|+(2\omega_
A-\alpha_A)|2\rangle_A\langle2|\notag\\
&&+\omega_B|1\rangle_\langle1|+(2\omega_B-\alpha_B)|2\rangle_B\langle2|]t\}
\end{eqnarray}
and
\begin{eqnarray}\label{B4}
V_2=\exp&&\{i[\mathrm{F}(t)|1\rangle_A\langle1|+2\mathrm{F}(t)|2\rangle_A\langle2|]\},
\end{eqnarray}
and the transformed Hamiltonian is
\begin{eqnarray}\label{B5}
H_t(t)&=&V^\dagger H_T(t)V+i\frac{dV\dagger}{dt}V\notag\\
&=&V^\dagger H_I(t)V \notag\\
&=&g[|0\rangle_A\langle1|e^{-i\omega_At}e^{i\mathrm{F}(t)}\notag\\
&&\quad +\sqrt{2}|1\rangle_A\langle2|e^{-i(\omega_A-\alpha_A)t}e^{i\mathrm{F}(t)}+\textrm{H.c.}] \\
&\otimes&[|0\rangle_B\langle1|e^{-i\omega_Bt}+
\sqrt{2}|1\rangle_B\langle2|e^{-i(\omega_B-\alpha_B)t}+\textrm{H.c.}],\notag
\end{eqnarray}
which leads to $H_t(t)$ in Eq. (\ref{Hint})  of the main text, after neglecting the high order
oscillating  terms.


\begin{thebibliography}{99}

\bibitem{GP1} M. V. Berry, Quantal phase factors accompanying adiabatic changes, Proc. R. Soc. Lond. Ser. A \textbf{392}, 45 (1984).

\bibitem{GP2} F. Wilczek and A. Zee, Appearance of gauge structure in simple dynamical systems, Phys. Rev. Lett.  \textbf{52}, 2111 (1984).

\bibitem{GP3} Y. Aharonov and J. Anandan, Phase change during a cyclic quantum evolution, Phys. Rev. Lett. \textbf{58}, 1593 (1987).

\bibitem{AN1} P. Solinas, P. Zanardi, and N. Zangh\`{\i}, Robustness of non-Abelian holonomic quantum gates against parametric noise, Phys. Rev. A \textbf{70}, 042316 (2004).

\bibitem{AN2} S.-L. Zhu and P. Zanardi, Geometric quantum gates that are robust against stochastic control errors, Phys. Rev. A \textbf{72}, 020301(R) (2005).

\bibitem{AN3} P. Solinas, M. Sassetti, T. Truini, and N. Zangh\`{\i}, On the stability of quantum holonomic gates, New J. Phys. \textbf{14}, 093006 (2012).

\bibitem{AN4} M. Johansson, E. Sj\"{o}qvist, L. M. Andersson, M. Ericsson, B. Hessmo, K. Singh, and D. M. Tong, Robustness of nonadiabatic holonomic gates, Phys. Rev. A \textbf{86}, 062322 (2012).

%
%


\bibitem{NA1} X. B. Wang and M. Keiji, Nonadiabatic conditional geometric phase shift with NMR, Phys. Rev. Lett. \textbf{87}, 097901
(2001)

\bibitem{NA2} S. L. Zhu and Z. D. Wang, Implementation of universal quantum gates based on nonadiabatic geometric phases, Phys. Rev. Lett. \textbf{89}, 097902 (2002).

\bibitem{NA3} P. Z. Zhao, X. D. Cui, G. F. Xu, E. Sj\"{o}qvist, and D. M. Tong, Rydberg-atom-based scheme of nonadiabatic geometric quantum computation, Phys. Rev. A \textbf{96}, 052316 (2017).

\bibitem{NA4} T. Chen and Z.-Y. Xue, Nonadiabatic geometric quantum computation with parametrically tunable coupling, Phys. Rev. Appl. \textbf{10}, 054051 (2018).

\bibitem{NA42} X.-Y. Chen, T. Li, and Z.-Q. Yin, Nonadiabatic dynamics and geometric phase of an ultrafast rotating electron spin, Sci. Bull. {\bf 64}, 380  (2019). 


\bibitem{NA5}  T. B{\ae}kkegaard, L. B. Kristensen, N. J. S. Loft, C. K. Andersen, D. Petrosyan, and N. T. Zinner,
Realization of efficient quantum gates with a superconducting qubit-qutrit circuit,
Sci. Rep. {\bf 9}, 13389 (2019).

\bibitem{NNA1} E. Sj\"{o}qvist, D. M. Tong, L. M. Andersson, B. Hessmo, M. Johansson, and K. Singh, Non-adiabatic holonomic quantum computation, New J. Phys. \textbf{14}, 103035 (2012).

\bibitem{NNA2} G. F. Xu, J. Zhang, D. M. Tong, E. Sj\"{o}qvist, and L. C. Kwek, Nonadiabatic holonomic quantum computation in decoherence-free subspaces, Phys. Rev. Lett. \textbf{109}, 170501 (2012).

\bibitem{exp1}
G. Falci, R. Fazio, G. M. Palma, J. Siewert, and V. Vedral,
Detection of geometric phases in superconducting nanocircuits,
Nature (London) \textbf{407}, 355 (2000).

\bibitem{exp2}
D. Leibfried, B. DeMarco, V. Meyer, D. Lucas, M. Barrett, J. Britton, W. M. Itano, B. Jelenkovi\'{c}, C. Langer, T. Rosenband, and D. J. Wineland,
Experimental demonstration of a robust, high-fidelity geometric two ion-qubit phase gate,
Nature (London) \textbf{422}, 412 (2003).


\bibitem{exp3}
P. J. Leek, J. M. Fink, A. Blais, R. Bianchetti, M. Goppl, J. M. Gambetta, D. I. Schuster, L. Frunzio, R. J. Schoelkopf, and A. Wallraff,
Observation of Berry's phase in a solid-state qubit,
Science \textbf{318}, 1889 (2007).

\bibitem{exp4} J.-M. Cui, M.-Z. Ai, R. He, Z.-H. Qian,  X.-K. Qin,  Y.-F. Huang, Z.-W. Zhou,  C.-F. Li,  T. Tu, and G.-C. Guo, Experimental demonstration of suppressing residual geometric dephasing, Sci. Bull. {\bf 64}, 1757 (2019). 



\bibitem{chu2019} J. Chu, D. Li, X. Yang, S. Song, Z. Han, Z. Yang, Y. Dong, W. Zheng, Z. Wang, X. Yu, D. Lan, X. Tan, and Y. Yu,
 Realization of Superadiabatic Two-qubit Gates Using Parametric Modulation in Superconducting Circuits,
Phys. Rev. Applied {\bf 13}, 064012 (2020).

\bibitem{xuy2019}  Y. Xu, Z. Hua, T. Chen, X. Pan, X. Li, J. Han, W. Cai, Y. Ma, H. Wang, Y. P. Song, Z.-Y. Xue, and L. Sun,
Experimental implementation of universal nonadiabatic geometric quantum gates in a superconducting circuit,
Phys. Rev. Lett. {\bf 124}, 230503 (2020).

\bibitem{NA6} P. Z. Zhao, Z. Dong, Z. Zhang, G. Guo, D. M. Tong, and Y. Yin, Experimental realization of nonadiabatic geometric gates with a superconducting Xmon qubit, arXiv, 1909.09970 (2019).


\bibitem{zheng}  S. B. Zheng, C. P. Yang, and F. Nori,
Comparison of the sensitivity to systematic errors between nonadiabatic non-Abelian geometric gates and their dynamical counterparts,
Phys. Rev. A {\bf 93}, 032313 (2016).

\bibitem{jing}  J. Jing, C.-H. Lam, and L.-A. Wu,
Non-Abelian holonomic transformation in the presence of classical noise,
Phys. Rev. A {\bf 95}, 012334 (2017).


\bibitem{ONN1} B.-J. Liu, X.-K. Song, Z.-Y. Xue, X. Wang, and M.-H. Yung, Plug-and-play approach to nonadiabatic geometric quantum gates, Phys. Rev. Lett. \textbf{123}, 100501 (2019).


\bibitem{ONN2} S. Li, T. Chen, and Z.-Y. Xue, Fast holonomic quantum computation on superconducting circuits with optimal control, Adv. Quantum Technol. {\bf 3}, 2000001 (2020).

\bibitem{yan2019}
T. Yan, B.-J. Liu, K. Xu, C. Song, S. Liu, Z. Zhang, H. Deng, Z. Yan, H. Rong, K. Huang, M.-H. Yung, Y. Chen, and D. Yu,   Experimental realization of nonadiabatic shortcut to non-Abelian geometric gates,
Phys. Rev. Lett. \textbf{122}, 080501 (2019).

 \bibitem{aimz2019} M.-Z. Ai, S. Li, Z. Hou, R. He, Z.-H. Qian, Z.-Y. Xue, J.-M. Cui, Y.-F. Huang, C.-F. Li, and G.-C. Guo, Experimental realization of nonadiabatic holonomic single-qubit quantum gates with optimal control in a trapped ion, arXiv, 2006.04609 (2020).

\bibitem{OC1} A. Ruschhaupt, X. Chen, D. Alonso, and J. G. Muga, Optimally robust shortcuts to population inversion in two-level quantum systems, New J. Phys. \textbf{14}, 093040 (2012).


\bibitem{OC2} D. Daems, A. Ruschhaupt, D. Sugny and S. Gu\'{e}rin, Robust quantum control by a single-shot shaped pulse, Phys. Rev. Lett. \textbf{111}, 050404 (2013).

\bibitem{OC3}   M. H. Goerz, F. Motzoi, K. B. Whaley, and  C. P. Koch, Charting the circuit QED design landscape using optimal control theory, npj Quantum Inf. {\bf 3},  37 (2017).

\bibitem{OC4}  G.  Bhole and J. A. Jones, Practical pulse engineering: Gradient ascent without matrix exponentiation, Front. Phys. {\bf 13}, 130312 (2018).


\bibitem{OC5} G. Long, G. Feng, and P. Sprenger, Overcoming synthesizer phase noise in quantum sensing, Quantum Engineering {\bf 1}, e27 (2019).

\bibitem{OC6}   K. Li, Eliminating the noise from quantum computing hardware, Quantum Engineering, {\bf 2}, e28 (2020).



\bibitem{SC1} J. Q. You and F. Nori, Atomic physics and quantum optics
using superconducting circuits, Nature \textbf{474}, 589 (2011).

\bibitem{SC2} M. H. Devoret and R. J. Schoelkopf, Superconducting circuits
for quantum information: An outlook, Science \textbf{339}, 1169 (2013).

\bibitem{SC3} X. Gu, A. F. Kockum, A. Miranowicz, Y.-X. Liu, and F. Nori, Microwave photonics with superconducting quantum
circuits,
Phys. Rep. {\bf 718}-{\bf 719}, 1 (2017).

\bibitem{SC4} M. Kjaergaard, M. E. Schwartz, J. Braum¨¹ller, P. Krantz, J. I.-J. Wang, S. Gustavsson, and W. D. Oliver,  Superconducting qubits: Current state of play, Annu. Rev. Condens. Matter Phys. \textbf{11}, 369 (2020). 

\bibitem{fp1}  Y.-J. Fan, Z.-F. Zheng, Y. Zhang, D.-M. Lu, and C.-P. Yang, One-step implementation of a multi-target-qubit controlled phase gate with cat-state qubits in circuit QED, Front. Phys. {\bf 14},  21602 (2019).

\bibitem{fp2}  X.-T. Mo and Z.-Y. Xue, Single-step multipartite entangled states generation from coupled circuit cavities, Front. Phys. {\bf 14}, 31602 (2019).

\bibitem{fp3}      H. Fan and X. Zhu, 12 superconducting qubits for quantum walks, Front. Phys. {\bf 14}, 61201 (2019).

\bibitem{DR1} J. M. Gambetta, F. Motzoi, S. T. Merkel, and F. K. Wilhelm, Analytic control methods for high-fidelity unitary operations in a weakly nonlinear oscillator, Phys. Rev. A \textbf{83}, 012308 (2011).

\bibitem{DR2} T. H. Wang, Z. X. Zhang, L. Xiang, Z. H. Gong, J. L. Wu, and Y. Yin,  Simulating a topological transition in a superconducting phase qubit by fast adiabatic trajectories, Sci. China Phys. Mech. Astro. \textbf{61}, 047411 (2018).

\bibitem{DR3} T. H. Wang, Z. X. Zhang, L. Xiang, Z. L. Jia, P. Duan, W. Z. Cai, Z. H. Gong, Z. W. Zong, M. M. Wu, J. L. Wu, L. Y. Sun, Y. Yin, and G. P. Guo, The experimental realization of high-fidelity 'shortcut-to-adiabaticity' quantum gates in a superconducting Xmon qubit, New J. Phys. \textbf{20}, 065003 (2018).





\bibitem{CP1}F. W. Strauch, P. R. Johnson, A. J. Dragt, C. J. Lobb, J. R. Anderson, and F. C. Wellstood, Quantum logic gates for coupled superconducting phase qubits, Phy. Rev. Lett. \textbf{91}, 167005 (2003).

\bibitem{CP2}L. DiCarlo, M. D. Reed, L. Sun, B. R. Johnson, J. M. Chow, J. M. Gambetta, L. Frunzio, S. M. Girvin, M. H. Devoret, and R. J. Schoelkopf, Preparation and measurement of three-qubit entanglement in a superconducting circuit, Nature \textbf{467}, 574 (2010).




\bibitem{CC1} M. Reagor \emph{et al}., Demonstration of universal
parametric entangling gates on a multi-qubit lattice, 
    Sci. Adv. \textbf{4}, eaao3603 (2018).

\bibitem{CC2} S. A. Caldwell \emph{et al}., Parametrically activated entangling gates using transmon qubits,  
    Phys. Rev. Appl. \textbf{10}, 034050 (2018).

\bibitem{CC3} X. Li, Y. Ma, J. Han, T. Chen, Y. Xu, W. Cai, H. Wang, Y. P. Song, Z.-Y. Xue, Z.-q. Yin, and L. Sun, Perfect quantum state transfer in a superconducting qubit chain with parametrically tunable couplings, Phys. Rev. Appl. \textbf{10}, 054009 (2018).





\end{thebibliography}
\end{document}